\documentclass[preprint,aps,prd,showpacs,nofootinbib]{revtex4}

\usepackage{graphicx}
\usepackage{amsfonts}
\usepackage{amsmath}

\newcommand{\cL}{\ensuremath{\mathcal{L}}}
\newcommand{\cM}{\ensuremath{\mathcal{M}}}
\newcommand{\cN}{\ensuremath{\mathcal{N}}}
\newcommand{\cO}{\ensuremath{\mathcal{O}}}

\newcommand{\cU}{\ensuremath{\mathcal{U}}}
\newcommand{\cV}{\ensuremath{\mathcal{V}}}
\newcommand{\cW}{\ensuremath{\mathcal{W}}}

\newcommand*{\chpt}{\raise0.4ex\hbox{$\chi$}PT}
\newcommand*{\schpt}{S\raise0.4ex\hbox{$\chi$}PT}

\newcommand*{\eg}{\textit{e.g.},\ }
\newcommand*{\etc}{\textit{etc.}}

\newcommand*{\Tr}{\operatorname{Tr}}

\newcommand{\ltwid}{\raise.3ex\hbox{$<$\kern-.75em\lower1ex\hbox{$\sim$}}}
\newcommand{\eq}[1]{Eq.~(\ref{eq:#1})}

\newcommand{\eqsthru}[2] {Eqs.~(\ref{eq:#1}) through (\ref{eq:#2})}

\begin{document}
\title{A Possible Aoki Phase for Staggered Fermions}
\author{C.\ Aubin}
\affiliation{Physics Dept., Columbia University, New York, NY 10027}
\author{Qinghai Wang}
\affiliation{Physics Dept., Washington University, St.\ Louis, MO 63130}
\begin{abstract}
  The phase diagram for staggered fermions is discussed in the context
  of the staggered chiral Lagrangian, extending previous work on the
  subject. When the discretization errors are significant, there 
  may be an
  Aoki-like phase for staggered fermions, where the remnant $SO(4)$
  taste symmetry is broken down to $SO(3)$. We solve explicitly for
  the mass spectrum in the 3-flavor degenerate mass case and discuss
  qualitatively the $2\!+\!1$-flavor case. From numerical results we
  find that current simulations are outside the staggered Aoki
  phase. As for near-future simulations with more improved versions of
  the staggered action, it seems unlikely that these will be in the
  Aoki phase for any realistic value of the quark mass, although the
  evidence is not conclusive.
\end{abstract}

\maketitle

\section{Introduction}\label{sec:intro}

Current simulations with dynamical staggered quarks have the
benefits of both good chiral properties at finite lattice spacing and
being computationally inexpensive \cite{Bernard:2002pd}. 
The price for these features is that the quark
doubling is not completely removed, and there are four ``tastes'' of
quarks for each staggered flavor. Additionally, there are large
$\cO(a^2)$ scaling violations that arise and must be taken into
account. Improved forms of the staggered action (\eg ``Asqtad''
staggered quarks) reduce these scaling violations but they are still
not negligible
\cite{Bernard:2002bk,Bernard:1999xx,Bernard:2000gd,Bernard:2001av,Aubin:2004wf}.

These scaling violations cause mass differences among the
various tastes of light pseudoscalar mesons 
that result from the four quark tastes.
In principle, these break the continuum $SU(4)$ taste
symmetry down to the lattice subgroup \cite{Golterman:1985dz},
however, at $\cO(a^2)$
there is a remnant $SO(4)$ taste symmetry which remains \cite{Lee:1999zx}. 
For each meson flavor, there are 15 tastes which 
fall into representations of $SO(4)$:
pseudoscalar (1), axial (4), tensor (6) and vector (4); 
there is also the $SU(4)$ singlet for a total of 16.
This approximate $SO(4)$ symmetry is broken down 
to the lattice subgroup at $\cO(a^4)$. These splittings may 
give rise to a negative mass-squared of a meson, since there is
no reason to assume the splittings are positive. 
This signals an instability of the vacuum, 
and thus a meson condensate may form. 

The idea that there could be this sort of spontaneous symmetry
breaking in the context of the flavor symmetry of Wilson fermions was
first proposed by Aoki \cite{Aoki:1983qi}. Lee and Sharpe showed that
this could also occur for one flavor 
of staggered fermions as well \cite{Lee:1999zx}. 
In this case, they assumed one of these splittings to be negative and
so the vector-taste 
pion condensed. All current formulations of staggered fermions 
find
that these mass splittings are all positive, and thus this is not a
likely scenario. In the case of multiple flavors, additional
parameters which mix the axial and vector taste flavor-neutrals
are negative in current simulations, so 
a possible ``staggered-Aoki'' phase may be more likely
\cite{Aubin:2003rg,Aubin:2003mg} than
originally thought. The phase
structure of lattice theories has been studied for a variety of
other lattice formulations recently
\cite{Sharpe:1998dw,Sharpe:1998xm,Munster:2004am,Sharpe:2004ps}.

As we approach the broken phase, 
the quark mass $m$ becomes comparable to the discretization errors;
in lattice units, $am \sim a^3\Lambda^3_{\rm QCD}$.  It is then that
$m^2_{\rm meson}$ may be negative and 
the $SO(4)$ taste symmetry is broken. The breaking is $SO(4)$ 
down to $SO(3)$, giving rise to three approximate
Goldstone bosons. We expect
a problem to arise in simulations if either very small quark
masses or large lattice spacings are used. It seems as though current
simulations are not in this region of the phase diagram, as
we will discuss. Although it is not completely 
ruled out, it appears as though
further improvements to the staggered action will also be safely away
from the staggered-Aoki phase.

As pointed out in Ref.~\cite{Lee:1999zx}, we are
not breaking an exact symmetry.
The $SO(4)$ symmetry will be broken
down to the lattice subgroup at higher order, so in this way it is
unlike the Aoki phase for Wilson fermions, where the flavor
symmetry is exact. Thus, the three Goldstone bosons actually have
masses of $\cO(a^4)$. Terms which arise at $\cO(a^4)$ are subleading,
however, and will only give rise to small corrections to our results
here. For small enough lattice spacing they cannot change drastically
the phase structure here. 

The goal of this paper is to study the features of the particle spectrum
in the broken phase as compared to the unbroken phase. We
will do this first for the case where all three quark masses are
degenerate, the 3-flavor case, which is 
simple and exactly solvable. We will then
look empirically at the $2\!+\!1$-flavor case, where the up and down
quark masses are degenerate but different than the strange quark
mass. This case is significantly more difficult than the 3-flavor 
case---we cannot analytically solve for the condensate in this case---and
we will simply make some general comments on the critical point and how
it relates to values of the parameters in current and future
simulations. We will find that current simulations are not in the
broken phase, and more
improved versions of staggered fermions are likely not going to be in
the broken phase, although we note that we cannot make exact
statements about the more improved versions without performing
simulations. 

The approach we follow in this paper is similar to that of 
Refs.~\cite{Sharpe:1998dw,Sharpe:1998xm,Munster:2004am,Sharpe:2004ps,Lee:1999zx},
using the chiral Lagrangian. 
This approach allows us to determine the mass spectrum 
of all the light pseudoscalar mesons rather simply. For the region
we are discussing, $m\sim a^2$, the chiral expansion is quite natural,
and we can use this to determine the meson masses with the dependence on
all parameters explicit. We are not performing a complete analysis of
the phase diagram, however. We will see that the phase diagram will
turn out to be more complex than studied in this paper, but, although
this will change certain features such as the mass spectrum, it will
not change some central results of the analysis, as we will see.

This paper is structured as follows. In Sec.~\ref{sec:stag_lag} we
layout the necessary tools for the analysis: 
the staggered chiral Lagrangian including the $\cO(a^2)$
taste-violating potential. Next, we determine the mass
spectrum for the case of 3 degenerate flavors of quarks in
Sec.~\ref{sec:mass_deg}.  In
Sec.~\ref{sec:mass_nondeg} we discuss the $2\!+\!1$-flavor case. Here
we will not solve for the mass spectrum, but make some qualitative
statements regarding this case. We will 
examine the features of the critical point with current simulations
as a function of the
quark masses in Sec.~\ref{sec:num}. 
Additionally we will look at how the point shifts
with more improved staggered fermion formulations, and will see
that indeed the problem becomes less severe in these
cases. We finish up with some conclusions in Sec.~\ref{sec:conc}.

\section{The Staggered Chiral Lagrangian}\label{sec:stag_lag}

The starting point of our analysis is the \schpt\
Lagrangian for 3 flavors of quarks \cite{Aubin:2003rg,Aubin:2003mg}.
The Lagrangian is written in terms of the field 
$\Sigma=\exp(i\Phi / f)$, a $12 \times 12$
matrix, with $\Phi$ given by:
\begin{eqnarray}\label{eq:Phi}
  \Phi = \left( \begin{array}{ccc}
    U  & \pi^+ & K^+  \\*
    \pi^- & D & K^0   \\*
    K^-  & \bar{K^0}  & S 
  \end{array} \right),
\end{eqnarray}
where the elements shown are each $4\times 4$ matrices,
linear combinations of the Hermitian generators
\begin{equation}\label{eq:T_a}
  T_a = \{ \xi_5, i\xi_{\mu5}, i\xi_{\mu\nu}, \xi_{\mu}, \xi_I\}.
\end{equation}
In other words, $U=\sum_a U_a T_a$, $K^0=\sum_a K^0_a T_a$,
\etc\ In Euclidean space, the gamma matrices $\xi_{\mu}$ are Hermitian, 
and we use the notations
$\xi_{\mu\nu}\equiv \xi_{\mu}\xi_{\nu}$ [$\mu <\nu$ in
\eq{T_a}], $\xi_{\mu5}\equiv \xi_{\mu}\xi_5$ and $\xi_I \equiv
I$ is the $4\times 4$ identity matrix. Under the chiral 
$SU(12)_L\times SU(12)_R$ symmetry, $\Sigma \to L\Sigma
R^{\dagger}$. The components of the diagonal (flavor-neutral)
elements ($U_a$, $D_a$, $S_a$) are real; while the other (flavor-charged)
fields are complex ($\pi^+_a$, $K^0_a$, \etc), such that $\Phi$ is
Hermitian. 

The Lagrangian is given by
\begin{eqnarray}\label{eq:final_L}
  \cL & = & \frac{f^2}{8} \Tr(\partial_{\mu}\Sigma 
  \partial_{\mu}\Sigma^{\dagger}) - 
  \frac{1}{4}\mu f^2 \Tr(\cM\Sigma+\cM\Sigma^{\dagger})
  + \frac{2m_0^2}{3}(U_I + D_I + S_I)^2 + a^2 \cV,
\end{eqnarray}
where $\mu$ is a constant with dimensions 
of mass, $f$ is the tree-level pion
decay constant, the $m_0^2$ term includes the 3 flavor-neutral taste-singlet
fields ($m_0\to\infty$ at the end to decouple the taste-singlet
$\eta'_I$), and
$\cV=\cU+\cU\,'$ is the taste-symmetry breaking potential given by
\begin{eqnarray}
  \label{eq:U}
  -\cU  & = & C_1
  \Tr(\xi^{(3)}_5\Sigma\xi^{(3)}_5\Sigma^{\dagger}) \nonumber \\*
  & & +C_3\frac{1}{2} \sum_{\nu}[ \Tr(\xi^{(3)}_{\nu}\Sigma
    \xi^{(3)}_{\nu}\Sigma) + h.c.] \nonumber \\*
  & & +C_4\frac{1}{2} \sum_{\nu}[ \Tr(\xi^{(3)}_{\nu 5}\Sigma
    \xi^{(3)}_{5\nu}\Sigma) + h.c.] \nonumber \\*
  & & +C_6\ \sum_{\mu<\nu} \Tr(\xi^{(3)}_{\mu\nu}\Sigma
  \xi^{(3)}_{\nu\mu}\Sigma^{\dagger}) \, ,\\
  \label{eq:U_prime}
  -\cU\,'  & = & C_{2V}\frac{1}{4} 
  \sum_{\nu}[ \Tr(\xi^{(3)}_{\nu}\Sigma)
    \Tr(\xi^{(3)}_{\nu}\Sigma)  + h.c.] \nonumber \\*
  &&+C_{2A}\frac{1}{4} \sum_{\nu}[ \Tr(\xi^{(3)}_{\nu
      5}\Sigma)\Tr(\xi^{(3)}_{5\nu}\Sigma)  + h.c.] \nonumber \\*
  & & +C_{5V}\frac{1}{2} \sum_{\nu}[ \Tr(\xi^{(3)}_{\nu}\Sigma)
    \Tr(\xi^{(3)}_{\nu}\Sigma^{\dagger})]\nonumber \\*
  & & +C_{5A}\frac{1}{2} \sum_{\nu}[ \Tr(\xi^{(3)}_{\nu5}\Sigma)
    \Tr(\xi^{(3)}_{5\nu}\Sigma^{\dagger}) ]\ .
\end{eqnarray}
The $\xi^{(3)}_B$ in
$\cV$ are block-diagonal $12\times 12$ matrices
\begin{equation}\label{eq:xi_B}
  \xi_B^{(3)} = 
  \left( \begin{array}{ccc}
    \xi_B & 0 & 0 \\*
    0 & \xi_B & 0 \\*
    0 & 0 & \xi_B 
  \end{array} \right),
\end{equation}
with $B \in \{5,\mu5,\mu\nu(\mu <\nu),\mu,I \}$.

The mass matrix, $\cM$, is the $12\times 12$ matrix
\begin{eqnarray}
  \cM = \left( \begin{array}{ccc}
    m_u I  & 0 &0   \\*
    0  & m_d I & 0  \\*
    0  & 0  & m_s I
  \end{array} \right) \ .
\end{eqnarray}
We will in later sections take either all three quark masses to be
degenerate or $m_u = m_d\ne m_s$. 

As noted in Refs.~\cite{Lee:1999zx,Aubin:2003mg}, 
this potential, although breaking the taste symmetry
at $\cO(a^2)$, has an accidental $SO(4)$ symmetry. This implies 
a degeneracy in the masses among different tastes of a given flavor
meson, 
which is seen in the
tree-level masses
of the pseudoscalar mesons. We can classify these mesons into irreducible
representations of $SO(4)$.
The mass for the meson $M$ (composed of quarks $a$
and $b$) with taste $B$, is given at tree-level by 
\begin{equation}\label{eq:tree_lev_mass}
  m^2_{M_B} = \mu \left(m_a + m_b \right) + 
  a^2\Delta_B,
\end{equation}
with
\begin{eqnarray}\label{eq:deltas}
  \Delta (\xi_5) & \equiv & \Delta_P  = 0
  \nonumber \\*
  \Delta (\xi_{\mu5}) & \equiv & \Delta_A = \frac{16}{f^2}\left( 
  C_1 + 3C_3 + C_4 + 3C_6 \right) \nonumber \\*
  \Delta (\xi_{\mu\nu})  & \equiv &\Delta_T =
  \frac{16}{f^2}\left(2C_3 + 2C_4 + 4C_6\right) \nonumber \\*
  \Delta (\xi_{\mu}) & \equiv & \Delta_V = \frac{16}{f^2}\left( 
  C_1 + C_3 + 3C_4 + 3C_6 \right) \nonumber \\*
  \Delta (\xi_I)  & \equiv & \Delta_I =
  \frac{16}{f^2}\left( 
  4C_3 + 4C_4 \right)\, .
\end{eqnarray}
As in Ref.~\cite{Aubin:2003mg}, we do not include the $m_0^2$ terms 
or the terms from $\cU\,'$
in these masses. Those terms only affect
the flavor-neutral mesons and allow for mixings 
in the basis of \eq{Phi}, the ``flavor basis.'' We can expand $\cU\,'$
to quadratic order to find these mixings take the same form as the
singlet term in the Lagrangian:
\begin{eqnarray}
  +\frac{a^2 \delta'_V}{2} (U_{\mu}+D_{\mu}+S_{\mu})^2 &\quad& 
  {\rm taste-vector} \\
  +\frac{a^2 \delta'_A}{2} (U_{\mu 5}+D_{\mu 5}+S_{\mu 5})^2 &\quad& 
  {\rm taste-axial} \, ,
\end{eqnarray}
where we have defined the parameters
\begin{equation}\label{eq:mix_vertex_V}
  \delta'_{V(A)} \equiv \frac{16}{f^2} \left[C_{2V(A)} - 
    C_{5V(A)}\right] \ .
\end{equation}
This 
requires us to resum the flavor-neutral propagators in the three cases:
taste-vector, taste-axial and taste-singlet 
\cite{Aubin:2003mg}. Performing this resummation
is equivalent to writing everything in terms of the 
``physical basis'' of our mesons, the $\pi^0$, $\eta$
and $\eta'$ for each of the three tastes here, which are the true
eigenstates of the mass matrix.
These are the additional parameters discussed in the introduction
which may cause the vector or axial $\eta'$ mass-squared 
to be negative, thereby introducing a staggered Aoki phase similar to
that discussed in Ref.~\cite{Lee:1999zx}. 

The mass eigenstates of the flavor-neutral mass matrix, 
for non-degenerate flavors, are quite complicated,
but the completely general case is not needed. The two cases
relevant to current simulations are the fully degenerate case 
($m_u = m_d = m_s \equiv m$), which
we refer to as the ``3-flavor'' case, and the case where two flavors
are degenerate ($m_u = m_d \equiv m \ne m_s$), or the ``$2\!+\!1$'' case. 
We will examine the 3-flavor case in Sec.~\ref{sec:mass_deg} and discuss the
$2\!+\!1$-flavor case in Sec.~\ref{sec:mass_nondeg}.

\section{Mass Spectrum in the Broken Phase for 
3 Degenerate Flavors}\label{sec:mass_deg}

We first discuss the simplest case, where all three quarks are
degenerate in mass, $m_u=m_d=m_s\equiv m$. First we need the 
eigenvalues of the full flavor-neutral mass matrix. 
The vector, axial and singlet tastes are the only tastes which mix
and therefore require rediagonalizing the mass matrix. The
singlet-taste 
masses do not involve the new parameters given in
\eq{mix_vertex_V}, and so we will not focus on their masses.
Also, for the singlet, we take $m_0\to\infty$ on theoretical grounds,
but there is no similar reasoning to dictate the 
values of $\delta'_{V}$ or $\delta'_A$; 
they must be determined from fits to lattice data
\cite{Aubin:2003ne,Aubin:2004fs}.

For the 3-flavor case, the mass
eigenvalues for the vector (axial) tastes are given by
\begin{eqnarray}\label{eq:mass_evalues_deg}
  m_{\pi^0_{V(A)}}^2 & = & m_{\eta_{V(A)}}^2 = 
  2\mu m + a^2\Delta_{V(A)}
  \nonumber\\
  m_{\eta'_{V(A)}}^2 & = &  2\mu m + a^2\Delta_{(V)A} +
  \frac{3}{4}a^2\delta'_{V(A)}\ .
\end{eqnarray}
We have taken into account the transition from four to one
tastes per flavor with the additional factor of $1/4$ in the
$a^2\delta'_{V(A)}$ term both in \eq{mass_evalues_deg} and below. 
More discussion on how to account for this
transition can be found in 
Refs.~\cite{Aubin:2003mg,Bernard:2001yj,Aubin:2003uc}. 

Lee and Sharpe's original discussion in the one-flavor case
focused on the possibility that
one of the splittings in \eq{deltas} may be negative and cause a
meson mass-squared to go negative for certain values of the quark
masses. This signals a vacuum instability and the 
formation of a meson condensate. As noted in Ref.~\cite{Lee:1999zx}, 
all current simulations with staggered quarks show that
$\Delta_B>0$ for all $B$ \cite{Bernard:2001av,Aubin:2004wf}. 
On the other hand, as is clear from
\eq{mass_evalues_deg}, there is another possibility for a meson
mass-squared to become negative, when one of the 
``hairpin'' parameters,
 $\delta'_{V(A)}$, is negative.\footnote{The term ``hairpin'' comes from the
fact that these two-point interactions are disconnected at the quark
level. See Refs.~\cite{Aubin:2003mg,Aubin:2003uc} for details.}
In current fits to MILC 
simulations, both $\delta'_{V}$ and $\delta'_{A}$ are negative
\cite{Aubin:2003ne,Aubin:2004fs}, so we can see it is more probable
to encounter an Aoki-like phase with staggered 
fermions in this way. Current
simulations are 
outside the broken phase
\cite{Bernard:2001av,Aubin:2004wf}. 
However, there is the question of how close current simulations are
to this broken phase and how improving the action more changes
this.

It is straightforward to see that the vector (axial) $\eta'$ mass-squared is
negative if
\begin{equation}\label{eq:crit_delta}
  a^2\delta'_{V(A)} < a^2\delta'_{V(A), \rm crit} 
  = -\frac{4}{3}\left[ 2\mu m
  + a^2\Delta_{V(A)}  \right]\ .
\end{equation}
$\Delta_A < \Delta_V$ in simulations, so \eq{crit_delta} implies that 
$|\delta'_{A, \rm crit}| < |\delta'_{V, \rm crit}|$. Thus, $\delta'_A$ is
more likely to be less than $\delta'_{A, \rm crit}$ than 
for the vector taste. We will assume for the following analysis that 
this is the case, although the analysis for the 
vector mesons would be analogous.

To find the vacuum state of the theory, 
we must minimize the potential
\begin{equation}
 \cW =  - \frac{1}{4}\mu m f^2 \Tr(\Sigma+\Sigma^{\dagger})
   + a^2 \cU + a^2 \cU\,' \ ,
\end{equation}
where we have substituted $\cM = {\rm diag}(mI,mI,mI)$.

This calculation is most simply done in the physical basis, where
everything is written in terms of $\pi^0$, $\eta$ and $\eta'$ instead
of the flavor-basis mesons $U$, $D$ and
$S$. In the 3-flavor
case, these two bases are related by
\begin{eqnarray}\label{eq:basis_transformation}
  \pi^0_B & = & \frac{1}{\sqrt{2}} \left( U_B - D_B  \right)\, ,  \nonumber\\
  \eta_B & = &  \frac{1}{\sqrt{6}} \left( U_B + D_B 
  - 2S_B \right)\, ,  \\
  \eta'_B & = & \frac{1}{\sqrt{3}} \left( U_B + D_B + S_B \right)\, ,\nonumber
\end{eqnarray}
with $B$ any of the 16 tastes.
In the $2\!+\!1$-flavor case, or any case where some masses are
non-degenerate, this relationship 
does not hold for all 
tastes.\footnote{We remark that often in continuum \chpt\ this
  relationship is used between the two bases. In that case, when
  $m_u\ne m_d$, there are additional mass-mixing terms between the
  $\pi^0$ and the $\eta$ which are usually neglected.} 
For the taste singlets, 
\eq{basis_transformation}
is valid in the $m_0\to\infty$ limit for any quark masses. 
For the tensor and pseudoscalar tastes, this also
holds.\footnote{We note that we need not write the tensor
and pseudoscalar taste mesons 
in the physical basis, since they do not mix. Since this
is an orthogonal transformation, 
the bases are equivalent.
Writing them in the physical basis makes the subsequent discussion
simpler.} For the vector and axial tastes, however, the relationship
is much more complicated in the non-degenerate case, although we will
not need the explicit relationship in that case.

Since we are assuming that the axial $\eta'$ mass-squared is going
negative, we are going to keep only this meson in $\Phi$.
This is valid right near the critical point and we assert
that this remains valid so long as no other mass-squared 
is negative in the new phase. If this occurs, then there will be a more
complicated pattern of symmetry breaking. We will see that this is a
possibility, and thus the phase diagram is richer than discussed
here. 
However, we will only focus on one region of the phase diagram
here. 

Keeping only the $\eta'_A$, $\Phi$ and $\Sigma$ are
block-diagonal in flavor space. We can
write the condensate $\langle\Sigma\rangle$ 
in terms of the five real numbers $A$ and $B_\mu$: 
\begin{equation}
  \langle\Sigma\rangle = A \left(1_{12\times 12}\right) + 
  	i B_\mu \left(i \xi^{(3)}_{\mu 5}\right)
\end{equation}
with the condition that $A^2 + B_\mu B_\mu = 1$. We write $A =
\cos\theta$ and substitute this into \cW\ to get (dropping constants)
\begin{equation}
  \cW = -3f^2 \left[
  2 \mu m \cos\theta 
  + \frac{1}{2} \left(a^2 \Delta_A + \frac{3}{4} a^2 \delta'_A
  \right)\cos^2\theta 
  \right]
\end{equation}
To find the condensate, we wish to minimize this potential. If
$\delta'_A$ is greater than $\delta'_{A, \rm crit}\equiv \delta_c'$,
then the minimum always lies at $\theta=0$ if $\mu m >0$ or
$\theta=\pi$ for $\mu m<0$.
In this phase, the unbroken phase, 
$\langle\Sigma\rangle = {\rm sign}(\mu m)1_{12\times 12}$
and the mass spectrum is
exactly that discussed above (for $\mu m <0$ make the replacement
$\mu m \to |\mu m|$ \cite{Lee:1999zx}).

The more interesting case is when we take $\delta'_A < 
\delta_c'$. Since $\delta'_A$ is negative in this region, we write
$\delta'_A = -|\delta'_A|$. Here, we find the potential is minimized when
\begin{equation}\label{eq:cond_sol}
  \cos\theta= \frac{8\mu m}{3a^2|\delta_A'| - 4a^2\Delta_A} \, ,
\end{equation}
so the 
condensate has the form
\begin{eqnarray}\label{eq:mass_matrix}
  \left \langle \Sigma   \right \rangle & =&
  \left( \begin{matrix}
     \exp \left[ i \theta
    n_\mu (i\xi_{\mu 5}) \right]& 0 &0  \nonumber\\*
    0 & \exp \left[ i \theta
    n_\mu (i\xi_{\mu 5}) \right] &  0 \nonumber \\*
    0 &0  & \exp \left[ i \theta
    n_\mu (i\xi_{\mu 5}) \right]
 \end{matrix}\right) \, ,
\end{eqnarray}
where $n_\mu$ is a unit vector.

The condensate breaks the
$SO(4)$ symmetry down to $SO(3)$, giving us three Goldstone bosons. As
noted in the introduction, these are not exactly Goldstone bosons,
because we are not breaking an exact symmetry. The $SO(4)$ 
symmetry will be broken down to the lattice subgroup
by terms of $\cO(a^4)$. Thus the masses of these pseudo-Goldstone Bosons are
$m^2_{\rm GB} \sim a^4\Lambda_{\rm QCD}^6$. 

Since the condensate is proportional to the identity 
in flavor space, the
flavor-charged mesons (the $\pi^+$, $K^0$, \etc) have the same
form as the $\pi^0$ and $\eta$
in the broken phase. Thus, below we only list the masses
for the $\pi^0$ of the various tastes, but keep in mind that 
$m^2_{\pi^0} = m^2_{\pi^+} =m^2_{\eta} =m^2_{K^+} =m^2_{K^0}$ for
all tastes.
Additionally, this form of the condensate implies that there is no
flavor
mixing between the $\pi^0$, $\eta$ and $\eta'$ in this case. We
pick the 4-direction for the condensate, so that $n_\mu =
\delta_{4\mu}$ to get explicit results for the masses. We also
define the quantities
\begin{eqnarray}
  \delta'_{2A} & = & \frac{16}{f^2}C_{2A} ,\\
  \delta'_{5A} & = & \frac{16}{f^2}C_{5A} ,
\end{eqnarray}
so $\delta'_A  = \delta'_{2A} - \delta'_{5A}$,
and similarly for $A\to V$.
We will organize the masses according to taste, 
by whether or not they mix.

\subsection{Tastes which mix}

The $\xi_4$ \& $\xi_5$ tastes mix with each other, as do the $\xi_{i4}$
\& $\xi_{i5}$ tastes (with $i=1,2,3$). For the $\xi_4$ \& $\xi_5$
tastes, we have the same mixing for any of the mesons.
With $\phi$ being $\pi^+$, $\pi^0$, $K^+$, $K^0$, 
$\eta$, or $\eta'$, we define the new fields
\begin{eqnarray}\label{eq:field_mixing}
  \phi_4^{\rm new} & = & \cos\theta \phi_4 - \sin\theta \phi_5 \\
  \phi_5^{\rm new} & = & \sin\theta \phi_4 + \cos\theta \phi_5 \ .
\end{eqnarray}
These new fields are defined so that at $\cos\theta=1$,
the critical point, $\phi_4^{\rm new}$ is 
purely $\phi_4$ and similarly $\phi_5^{\rm new} = \phi_5$. 
This means that in the chiral limit, at
$\cos\theta=0$, this is reversed: $\phi_4^{\rm new}$ is 
purely $\phi_5$ and vice versa.
In terms of these new mass eigenstates, we find the masses
\begin{eqnarray}
  m^2_{\pi^{0, \rm new}_4}  
  & = & a^2 \Delta_V - a^2 \Delta_A + \frac{3}{4} a^2 |\delta_A'|
  \label{eq:newpi4}\\
  m^2_{\pi^{0, \rm new}_5}  
  & = & m^2_{\eta'_5{}^{\rm new}}  
  =-  a^2 \Delta_A + \frac{3}{4} a^2 |\delta_A'| \\
  m^2_{\eta'_4{}^{\rm new}}
  & = &  a^2 \Delta_V - a^2 \Delta_A +
  \frac{3}{4}a^2  |\delta_A'| - \frac{3}{4}a^2
  \delta_V'\label{eq:newetap4}
\end{eqnarray}
Note 
the combination $-4a^2 \Delta_A + 3 a^2 |\delta_A'|$,
which appears frequently, is positive, so
\eqsthru{newpi4}{newetap4} are all positive in the broken
phase. 

As for the $\xi_{i4}$ \& $\xi_{i5}$ tastes, we have the same mixing as
in \eq{field_mixing}, with $4\to i4$ and $5\to i5$, again with $\phi$
referring to any of the six mesons.
These have the masses
\begin{eqnarray}
  m^2_{\pi_{i4}^{0, \rm new}} 
  & = & m^2_{\eta'_{i4}{}^{\rm new}} = 
  a^2\Delta_T - a^2 \Delta_A + \frac{3}{4}a^2  |\delta_A'| \\
  m^2_{\pi_{i5}^{0, \rm new}}   & = & 
  \frac{3}{4} a^2 |\delta_A'| \\
  m^2_{\eta'_{i5}{}^{\rm new}} & = & 0\, .
\end{eqnarray}
Again, all of these squared 
masses are positive for our range of parameter values.
We see that the 
$\eta'_{i5}{}^{\rm new}$ are the three expected Goldstone bosons. 

\subsection{Tastes which do not mix}

The rest of the tastes do not mix
with each other, and so we just list the masses for each of these states.
\begin{eqnarray}
  m^2_{\pi^0_I} & =&
  a^2 \Delta_V-a^2 \Delta_A + \frac{3}{4}a^2 |\delta_A'|
  + \frac{64 m^2 \mu^2\left(a^2\Delta_I-a^2\Delta_V\right)}
  {\left(4a^2\Delta_A-3a^2|\delta_A'|\right)^2}\\
  m^2_{\pi^0_{45}}& =& 
  - a^2 \Delta_A + \frac{3}{4}a^2 |\delta_A'|
  + \frac{64 m^2 \mu^2 a^2 \Delta_A}
    {\left(4a^2\Delta_A-3a^2|\delta_A'|\right)^2}\\
  m^2_{\eta'_{45}} &=& - a^2 \Delta_A 
  + \frac{3}{4}a^2 |\delta_A'|
  + \frac{16 m^2 \mu^2}{4a^2\Delta_A - 3a^2|\delta_A'| }\\
  m^2_{\pi^0_{i}} &= &
  a^2 \Delta_T-a^2 \Delta_A + \frac{3}{4}a^2 |\delta_A'|
  + \frac{64 m^2 \mu^2\left(a^2\Delta_V-a^2\Delta_T\right)}
  {\left(4a^2\Delta_A - 3a^2|\delta_A'| \right)^2}\\
  m^2_{\eta'_{i}} 
 & = &  a^2 \Delta_T-a^2\Delta_A + \frac{3}{4}a^2|\delta_A'|
  + \frac{16 m^2 \mu^2\left(4a^2\Delta_V - 4a^2\Delta_T +  
    3a^2\delta_V'\right)}
  {\left(4a^2\Delta_A - 3a^2|\delta_A'| \right)^2}\\
  m^2_{\pi^0_{ij}} &=& \frac{3}{4} a^2|\delta_A'| + 
  \frac{64 m^2 \mu^2\left(a^2\Delta_T-a^2\Delta_A \right)}
       {\left(4a^2\Delta_A - 3a^2|\delta_A'| \right)^2} \\
  m^2_{\eta'_{ij}} &= &
  \frac{3}{4}a^2|\delta_A'| - \frac{3}{4}a^2\delta_{2V}'
  -\frac{3}{4}a^2\delta_{5V}' \nonumber\\&&{}+ 
  \frac{16 m^2 \mu^2\left(4a^2\Delta_T - 4a^2\Delta_A + 
    3 a^2\delta_{2V}' + 3a^2\delta_{5V}'\right )}
       {\left(4a^2\Delta_A - 3a^2|\delta_A'| \right)^2}.
\end{eqnarray}

In the chiral limit, when $m = 0$, we have several degeneracies,
similar to those found in Ref.~\cite{Lee:1999zx}. In this limit, the
states which mix are purely one state (\eg 
$\pi_{i4}^{0,\rm new} = \pi_{i5}^{0}$), and we denote these by the
pure state. When $m=0$, the $U(1)_A$ is restored, and we see that 
$m^2_{\pi_{i5}^0} = m^2_{\pi^0_i} = m^2_{\eta'_i}$, 
$m^2_{\pi_4^0} = m^2_{\pi^0_{45}} = m^2_{\eta'_{45}}$ and
$m^2_{\pi_5^{0}} = m^2_{\pi^0_I}$. 
Additionally, we have
$m^2_{\pi_{i4}^0} = m^2_{\pi^0_{ij}}$, which occurs in the massless
case due to an accidental $SO(3)$ symmetry about the minimum of the
potential when $\delta'_A=0$. 
This degeneracy recurs at 
the critical point for a different
reason: due to the restoration of the $SO(4)$ symmetry.
Note that the limits of $m\to 0$ and
$\delta'_A\to\delta'_c$ do not commute. If we take
$\delta'_A\to\delta'_c$ first, the masses all become those of the
unbroken phase, so taking $m\to 0$ gives the massless forms of
\eq{tree_lev_mass}. 
However, taking $m\to 0$ first while in the broken phase then
the $\delta'_A\to\delta'_c$ limit results in a much different spectrum
which is not the same as in the unbroken phase.

The appearance
of the ``incomplete'' hairpin parameters
$\delta_{2V}'$ and $\delta_{5V}'$ in the $\eta'_{ij}$ mass shows that
the phase structure is 
more complicated than just the two phases we have discussed. Since we
do not know the size or the sign of either of these two parameters,
these could cause the mass-squared of the $\eta'_{ij}$ to go
negative. If this is the case, the spectrum we have laid out will
not be strictly 
valid, and we would have to take into account the possibility of
a condensate also in the $ij$-directions. 

We can determine 
the requirements on the parameters to cause the 
$\eta'_{ij}$ mass-squared to be negative. This
happens when 
\begin{equation}\label{eq:incomplete}
a^2\delta'_{5V} + a^2\delta'_{2V} > \frac{
  3a^2|\delta'_A| (3a^2|\delta'_A| - 4a^2\Delta_A)^2 +
  256 (a^2\Delta_T - a^2\Delta_A) \mu^2 m^2 }
  {3 (3 a^2|\delta'_A| - 4a^2 \Delta_A)^2 - 
    192 \mu^2 m^2} \, .
\end{equation}
So in the massless case (where we will see later that we are most
likely to be in the broken phase),
this condition requires 
$a^2\delta'_{5V} + a^2\delta'_{2V} > a^2|\delta'_A|$.
So the 
further we go into the broken phase, $|\delta'_A|$ gets larger,
meaning that we have a 
less chance of finding 
$m^2_{\eta'_{ij}} < 0$. Here we are studying the
simplest case, where this criterion is not met, and the spectrum above
is valid.

We cannot know the values of these parameters
from first principles; we must calculate them from fits
to lattice data. This
can be done by calculating quantities for non-Goldstone
mesons (such as the taste-singlet or taste-vector charged pions). 
The specific combination $\delta'_A = \delta'_{2A} -
\delta'_{5A}$ (and similarly for $A\to V$) 
appears when calculating the pseudoscalar taste masses and decay
constants, but in other taste mesons, we find that $\delta'_{2A} +
\delta'_{5A}$ also appears.
In order to have a complete determination of the
taste-violating parameters, one must calculate the expressions for the
masses and decay constants for each of the other tastes and match
those expressions to lattice simulation data.

This will not change our results
in Sec.~\ref{sec:num}, however, since those results are
focused on studying the
location of the critical point, $\delta'_c$. The criterion 
$\delta'_A>\delta'_c$ is required for any symmetry breaking; 
if this does not hold we have the mass spectrum of the
unbroken phase. If \eq{incomplete} holds, this would change the
spectrum of masses and the breaking pattern as well as the condensate
itself.  
We will discuss the functional dependence of
$\delta'_c$ on the various parameters, and examine the possibility of
current (and near future) simulations being performed in the broken phase.
If we are in the broken phase, then we would have to
determine if \eq{incomplete} holds; if not, the spectrum discussed
above would be correct. If it does, a more complete study should be
done of the resulting mass spectrum. As it is, a
very rich phase structure 
is possible
for staggered fermions. 

\section{The Broken Phase for 
$2\!+\!1$ Flavors}\label{sec:mass_nondeg}
In the $2\!+\!1$ case, we set $m_u = m_d \equiv m_l \ne m_s$, 
so $\cM = {\rm diag}(m_lI, m_lI, m_sI)$. 
Here when diagonalizing
the flavor neutral axial taste mass matrix, the mass
eigenstates for the axial tastes to be given by \cite{Aubin:2003mg}
\begin{eqnarray}\label{eq:eigenvalues_A}
  m_{\pi^0_A}^2 & = & 2\mu m_l + a^2 \Delta_A \ ,   \nonumber \\*
  m_{\eta_A}^2 & = & 
  \frac{1}{2}
  \left[ 2\mu (m_l + m_s) + 2a^2\Delta_A 
  \frac{3}{4}a^2\delta'_A + Z
  \right]\ , \nonumber \\*
  m_{\eta'_A}^2 & = &  \frac{1}{2}\left[ 2\mu (m_l + m_s) + 2a^2\Delta_A +
  \frac{3}{4}a^2\delta'_A -  Z \right] \ ; \\
  Z & \equiv &\sqrt{ 4 \mu^2\left( m_s - m_l \right)^2
    - a^2\delta'_A\mu 
    \left(m_s - m_l\right) +\frac{9(a^2\delta'_A)^2}{16}
  } \label{eq:Z}  \, .
\end{eqnarray}
The vector tastes have the same form (with $V\to A$), but again
we are only focusing on the axial taste masses vanishing.

Our definitions of $\eta_A$ and
$\eta'_A$ are different from that in 
Ref.~\cite{Aubin:2003mg}. We define the
$\eta_A$ as the field which becomes degenerate with the $\pi^0_A$ when
$m_l=m_s$. Thus the $\eta_A$ from Ref.~\cite{Aubin:2003mg} 
is now our $\eta'_A$ since
$\delta'_A<0$ (as determined from simulations). 
Using this definition, it is the $\eta'_A$ mass which vanishes
at the critical value of $\delta'_A$, just as in the 3-flavor case.
In the $2\!+\!1$-flavor case, this critical point is given by the expression
\begin{equation}\label{eq:crit_delta_21}
  a^2\delta'_c = -\frac{4\left(2\mu m_l + a^2\Delta_A \right)
    \left(2\mu m_s + a^2\Delta_A\right)}
  {2\mu \left(m_l+2m_s\right) + 3a^2\Delta_A}
  \, .
\end{equation}

The condensate has
the form
\begin{eqnarray}\label{eq:cond_2plus1}
  \left \langle \Sigma   \right \rangle & = & 
  \left(   \begin{matrix}
   \exp \left[ i \theta n_\mu (i\xi_{\mu 5}) \right]  & 0 & 0\\
    0 & \exp \left[ i \theta n_\mu (i\xi_{\mu 5}) \right] & 0 \\
    0 & 0 &\exp \left[ i \kappa\theta n_\mu (i\xi_{\mu 5}) \right]  
	   \end{matrix}\right)\, ,
\end{eqnarray}
$n_\mu$ is a unit vector in the $\mu$ direction and we have defined
\begin{equation}
  \kappa = \frac{8\mu (m_s - m_l) - a^2\delta'_A 
    - 4Z}{2a^2\delta'_A}\ ,
\end{equation}
which arises from the relationship between the $\eta'_A$ and the
flavor basis
\begin{equation}
  \eta'_A \equiv \cN \left( U_A + D_A + \kappa S_A  \right)\ ,
\end{equation}
with $\cN$ a normalization constant. Note that in the limit that
$\delta'_A\to -\infty$, or $m_s\to m_l$, $\kappa\to 1$, giving us the
same definition of the $\eta'_A$ as in the 3-flavor case.

$\theta$ in \eq{cond_2plus1} is the solution which minimizes the
potential, which in this case has the form
\begin{eqnarray}\label{eq:2p1_pot}
  \cW & = & -f^2\Biggl[
    4\mu m_l \cos(\theta) + 2\mu m_s \cos(\kappa\theta)
    + \left( a^2\Delta_A + \frac{1}{2}a^2\delta'_A \right)\cos^2(\theta)
    \nonumber\\*
    &&{}+ \frac{1}{2}\left( 
    a^2\Delta_A + \frac{1}{4}a^2\delta'_A \right)\cos^2(\kappa\theta)
    - \frac{1}{2}a^2\delta'_A\sin(\theta)\sin(\kappa\theta)
    \Biggr] + {\rm const.}
\end{eqnarray}

\begin{figure}
 \includegraphics[width=5in]{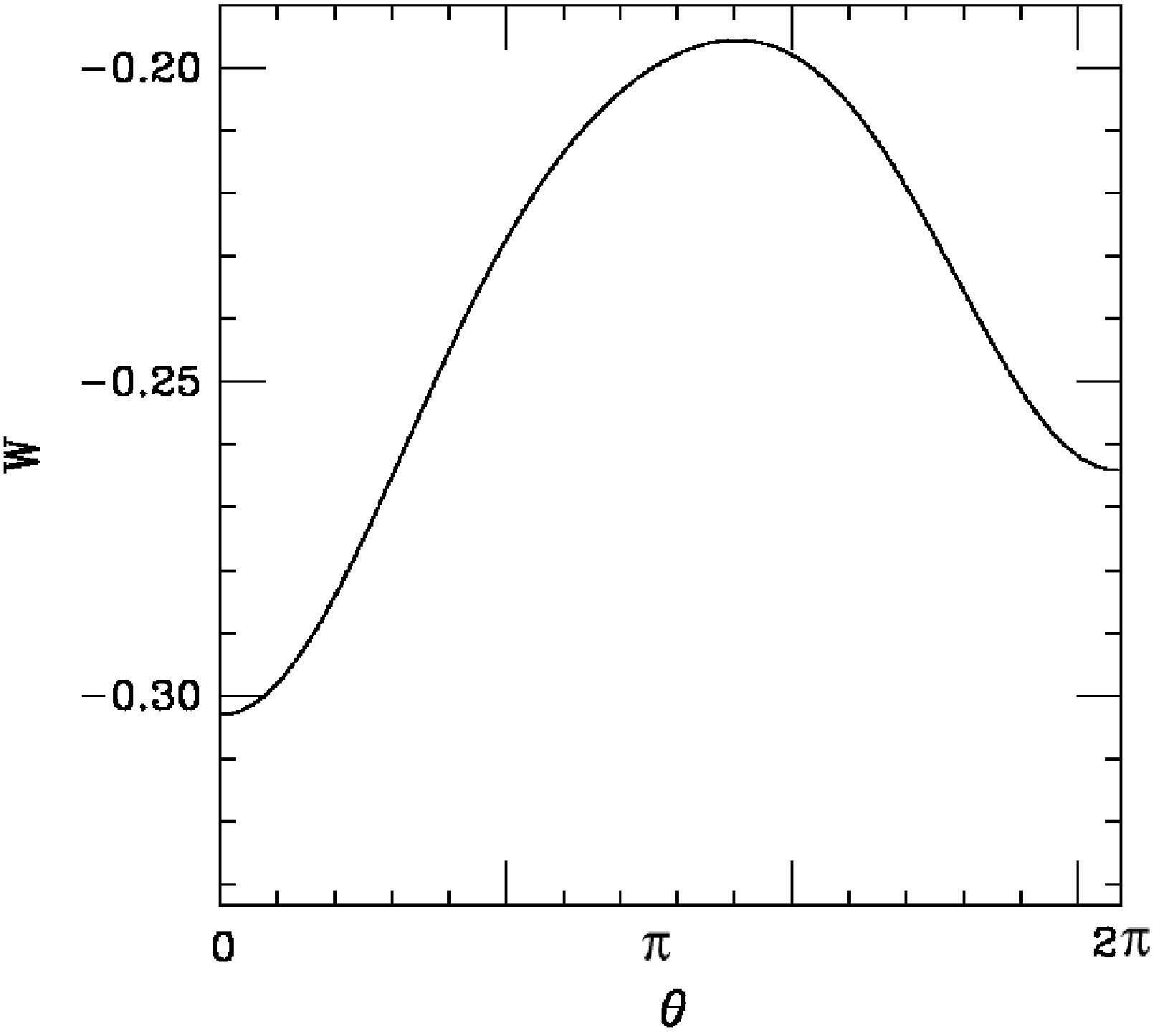} 
 \caption{Potential in the unbroken phase; parameters here are 
   for the coarse MILC lattice, with the light/strange lattice
   quark masses
   $0.005/0.05$.
   The global minimum is clearly at $\theta=0$.
   }
 \label{fig:pot_unbroken}
\end{figure}

\begin{figure}
 \includegraphics[width=5in]{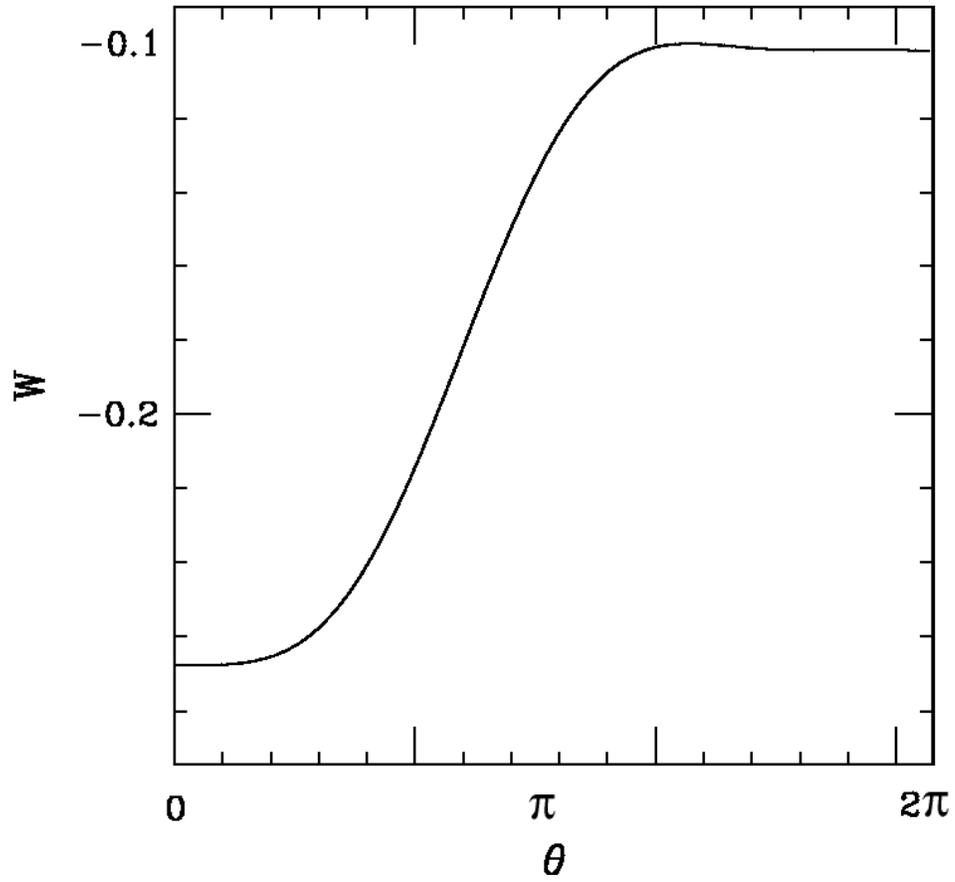} 
 \caption{The potential at the critical point, for the same masses as
   in Fig.~\ref{fig:pot_unbroken}. The region near
   $\theta=0$ has become flatter, as expected near a critical point.
   }
 \label{fig:pot_crit_point}
\end{figure}

\begin{figure}
 \includegraphics[width=5in]{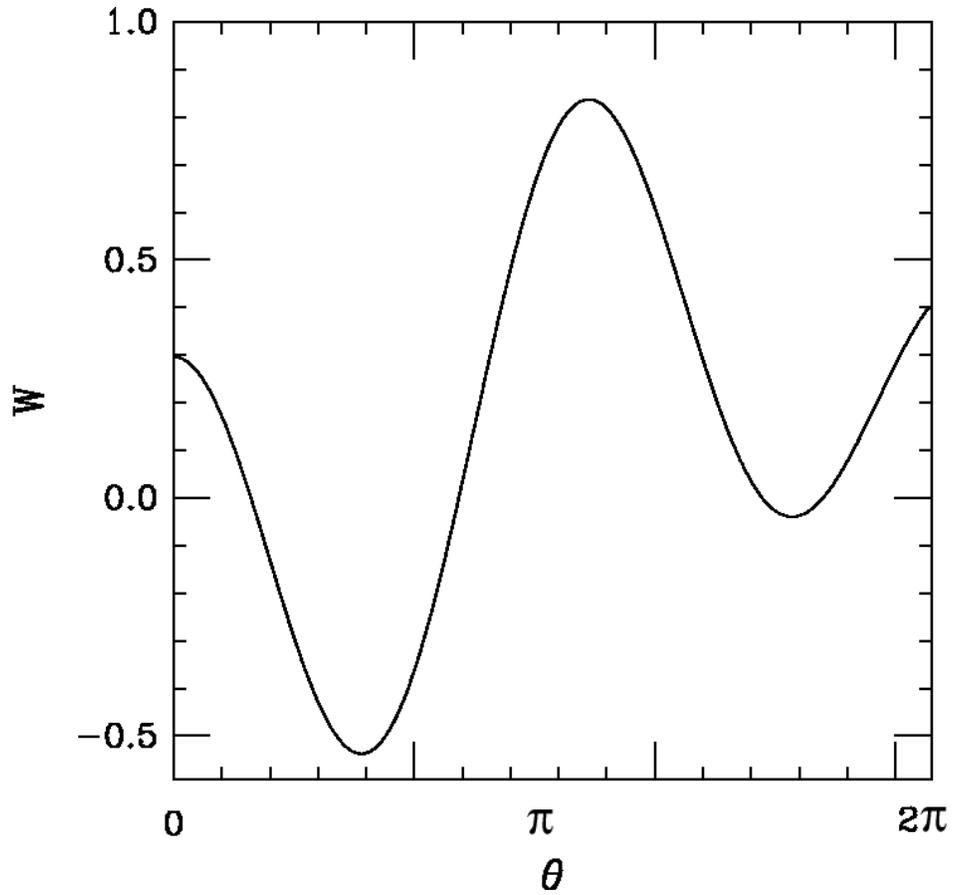} 
 \caption{The potential in the broken phase, where the global minimum
   now is at a non-zero value of $\theta$. Note the other (local)
   minimum has shifted down and to the left; 
   as $\delta'_A\to - \infty$, the
   two minima will both be global minima at different values of
   $\theta$.
   }
 \label{fig:pot_broken}
\end{figure}

We plot this potential in Fig.~\ref{fig:pot_unbroken} for the unbroken
phase (so that $\delta'_A>\delta'_c$). In this and subsequent figures,
we are plotting the dimensionless quantity $W = a^2\cW / f^2$ as a
function of $\theta$.  There are two minima, one of which is the
global minimum at $\theta=0$, which defines the vacuum state as we
expect in the unbroken phase. At the critical point, $\delta'_A =
\delta'_c$, shown in Fig.~\ref{fig:pot_crit_point}, both minima have
flattened out; if we were to expand \eq{2p1_pot} about $\theta=0$, we
would see the $\theta^2$ term vanish here. 
Once we have entered the broken phase, in
Fig.~\ref{fig:pot_broken}, the minima have both shifted towards each
other, and the global minimum is at a non-zero value of $\theta$. As
$\delta'_A\to -\infty$, we find that the minima correspond to two
equivalent solutions for the ground state here.  
In terms of the potential and the critical point, the $\delta'_A\to
-\infty$ limit is similar to the chiral limit [see \eq{crit_delta_21}], 
where the $U(1)_A$ symmetry is restored. In the chiral limit,
there are two minima given by $\theta=0$ or $\pi$, 
and they are equivalent.
Shifting $\theta$
by $\pi$ merely sends $\Sigma\to-\Sigma$, which is a symmetry of the
Lagrangian in the massless case. In the case of $\delta'_A\to
-\infty$, the minima are not at $0$ or $\pi$, but roughly near $\pi/2$
and $3\pi/2$, and are equivalent minima in the potential.

We cannot solve for an analytic
solution for $\cos\theta$ in this case as we did in the 3-flavor case,
although we can solve for the masses of the
mesons in the broken phase in terms of $\theta$, or more appropriately
$\cos(\theta)$ and $\cos(\kappa\theta)$.  
As in the 3-flavor case, there is taste mixing, but now the condensate
is not proportional to the identity in flavor space. 
This implies that we should find
mixing between the
$\eta$ and $\eta'$ mesons. Since the condensate is still diagonal in
flavor space, there will be no mixing between the charged fields.
The up and down quarks are
degenerate, so the $\pi^0$, $\pi^+$ and $\pi^-$ have 
the same general form; there
are only mixings among the tastes but there is no flavor mixing.  The
expressions are complicated and not very enlightening, so instead of
looking directly at the mass spectrum, we will study this phase
numerically. In the next section, we will look at the critical point
as a function of the various parameters and see the criteria for
entering this staggered Aoki phase.

\section{Numerical results}\label{sec:num}

With current MILC simulations, we now discuss the 
dependence of the critical point on the various parameters
in our theory. We will look at data from the 
``coarse'' ($a\approx 0.125$fm) and ``fine'' ($a\approx
0.09$fm) MILC lattices \cite{Bernard:2001av,Aubin:2004wf}. 
We will look first at the quark mass dependence on the two
different lattices with all other parameters fixed, and
then see what happens at fixed quark mass when we decrease
the size of the splitting, $\Delta_A$. This would correspond 
to staggered actions that have smaller discretization 
errors than the Asqtad action, such as HYP staggered fermions
\cite{Hasenfratz:2001hp,Hasenfratz:2002vv}.

\begin{figure}
 \includegraphics[width=5in]{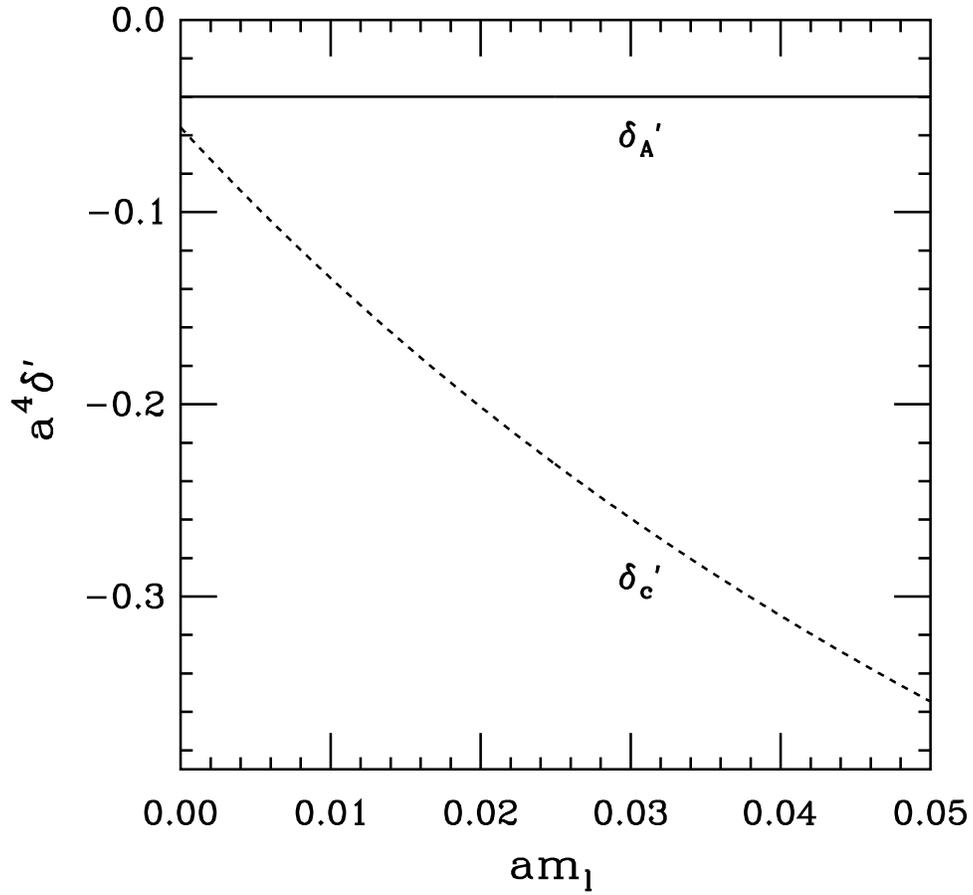} 
 \caption{The dashed line shows $\delta'_c$ as a function
   of the light quark mass (in lattice units) on the coarse 
   lattice, for a lattice strange
   quark mass of $0.05$. The solid horizontal line is the value of
   $\delta'_A$, 
   determined from chiral fits to the $2\!+\!1$ MILC data 
   \cite{Aubin:2003ne,Aubin:2004fs}.
   }
 \label{fig:delta_vs_mass_c}
\end{figure}

\begin{figure}
 \includegraphics[width=5in]{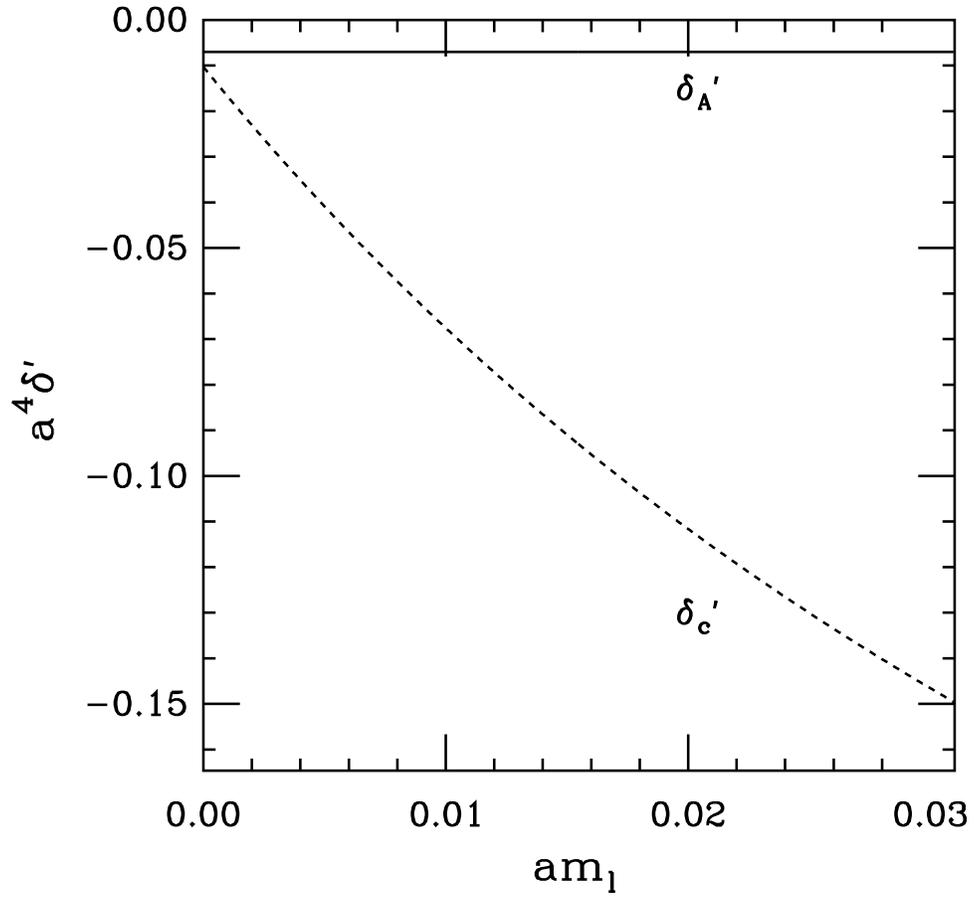} 
 \caption{Same as Fig.~\ref{fig:delta_vs_mass_c} but for the 
 	 fine lattice spacing, and for a lattice strange
	 quark mass of $0.031$.  }
 \label{fig:delta_vs_mass_f}
\end{figure}

Figures~\ref{fig:delta_vs_mass_c} and \ref{fig:delta_vs_mass_f}
show the critical $\delta'_A$, 
\eq{crit_delta_21}, on the coarse and fine
lattices, respectively, as a function of
light quark mass at fixed strange quark mass. 
In lattice units, the strange quark mass is 
$0.05 (0.031)$ on the coarse (fine)
lattice. As noted in Ref.~\cite{Aubin:2004wf}, 
these values are slightly larger than the
physical strange quark mass, but there is no qualitative difference in the
results here when we lower $am_s$ by $\approx 20\%$.
The horizontal line shows the value of $\delta'_A$ which was
determined in chiral fits to the mass and decay constant of the
pseudoscalar-taste mesons \cite{Aubin:2003ne,Aubin:2004fs}. 
We can see that we are far outside the
broken phase here, for both the coarse and the fine lattices, even
as we take the chiral limit. The relevant points in the parameter
space are when $am_l \ge 0.005(0.0062)$, since this is the range of
values simulated currently on the coarse (fine) MILC lattices, and
$am_l \approx 0.002(0.001)$, the physical value of the average up and
down quark masses on the coarse (fine) MILC lattices.\footnote{These
  values for the physical light quark mass are found relative to the
  ``nominal'' strange quark mass, 
  $am_s = 0.05(0.031)$ on the 
  coarse(fine) lattices, using the ratio $m_s/m_l
  \approx 27$ \cite{Aubin:2004ck}.}

\begin{figure}
 \includegraphics[width=5in]{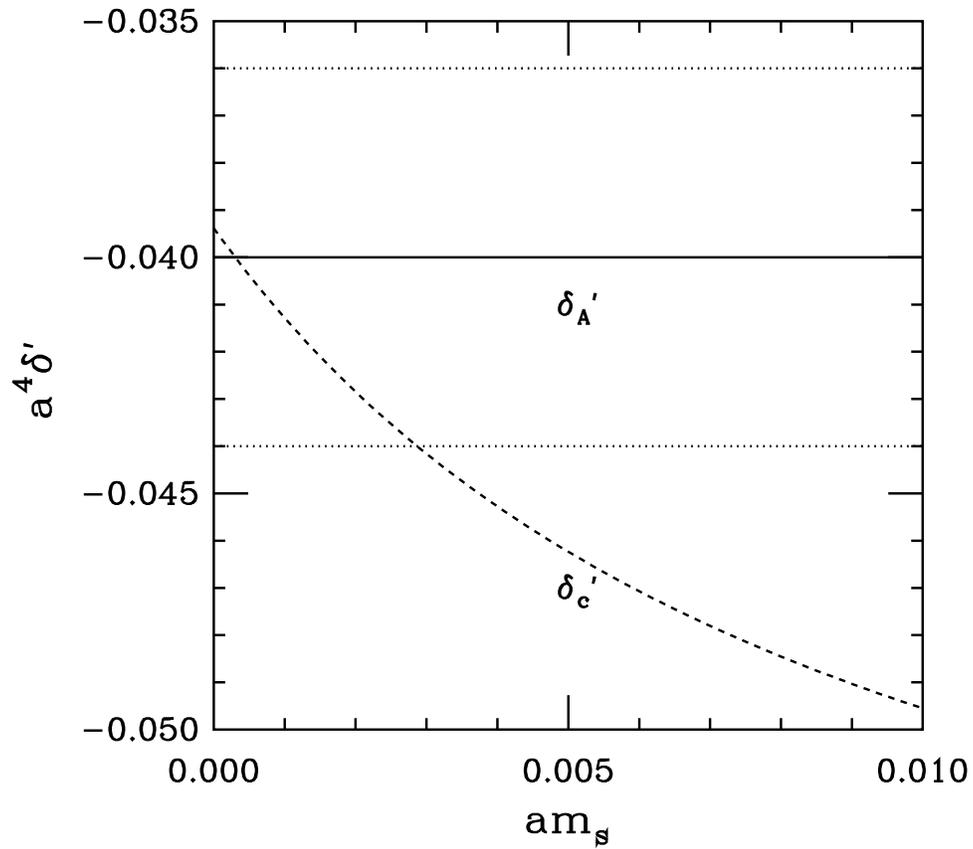} 
 \caption{The dashed line is $\delta'_c$ on the coarse
   lattice as a function of the strange quark mass for $m_l=0$. The solid line is the
   value of $\delta'_A$, with the statistical errors given by the
   dotted lines. For very light $m_s$ we see that 
   there is a significant probability that we will enter the broken phase.}
 \label{fig:delta_vs_ms_c}
\end{figure}

\begin{figure}
 \includegraphics[width=5in]{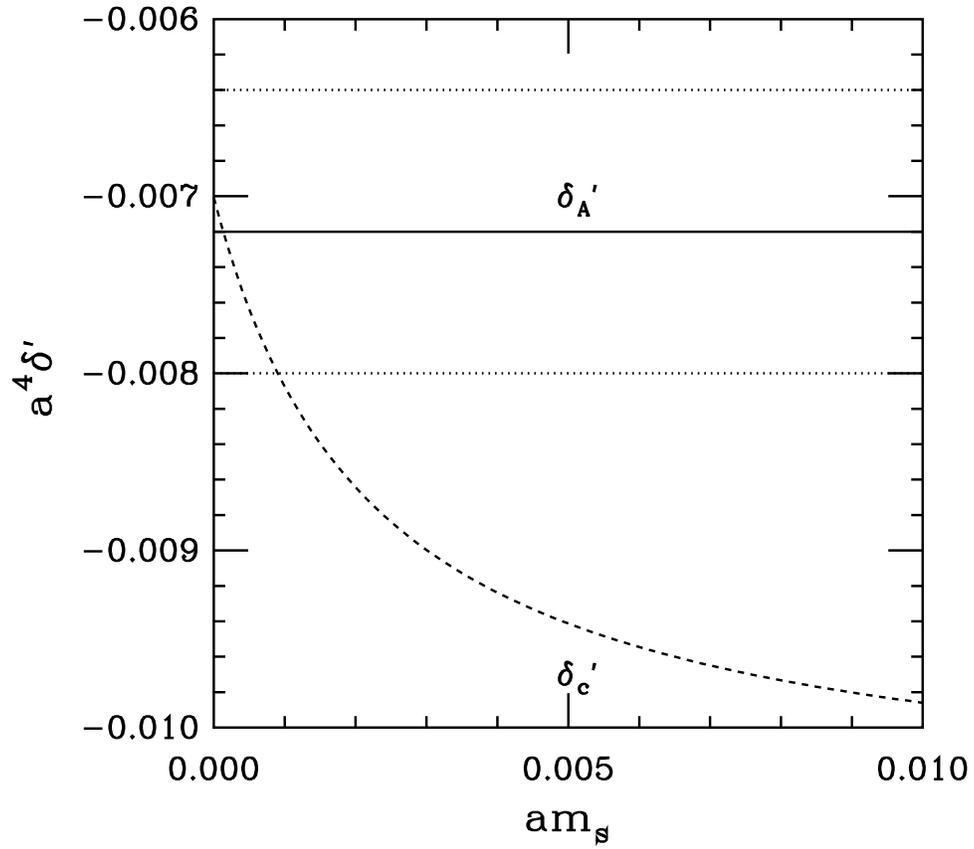} 
 \caption{Same as Fig.~\ref{fig:delta_vs_ms_c} but for the 
   fine lattice spacing. 
   There is still a significant probability that we could be in the broken phase for light 
   $m_s$ but it is somewhat less likely than on the coarse latice.}
 \label{fig:delta_vs_ms_f}
\end{figure}

We move closer to the broken phase when all three quark masses are
light: the 3-flavor case in the chiral limit. In this case,
$\delta'_c = -4\Delta_A/3$, which is 
surprisingly close to the value of $\delta'_A$ currently determined
from simulations \cite{Aubin:2003mg}. Figures~\ref{fig:delta_vs_ms_c} 
and \ref{fig:delta_vs_ms_f} shows $\delta'_c$ as a function of the
strange quark mass at $m_l=0$ for the coarse and fine lattices 
respectively. The solid line shows the central value of $\delta'_A$
that has been determined, while the dotted lines give the statistical
error. 
We can see that there is a better 
chance to enter the broken phase for very light strange quark
mass, although the error is large enough that this may not
occur. A more precise determination of the hairpin parameter would
allow for a concrete statement 
about the broken phase at light quark masses.

As is apparent from Figs.~\ref{fig:delta_vs_ms_c} 
and \ref{fig:delta_vs_ms_f}, as $a$ becomes smaller, we are
somewhat less likely 
to be in the broken phase. 
This seems to imply that as $a\to 0$, there is little chance we
will be in the broken phase, even close to the chiral limit. This
should be expected, as this phase is a lattice artifact and thus
vanishes in the continuum limit (note that as $a\to 0$,
$\delta'_c\to -\infty$). This also
makes it clear that at larger lattice spacing, $\delta'_A$ is more
likely to be in this phase. We anticipated this, since the broken
phase appears when the discretization errors are comparable to the
lattice quark masses. This implies that we must not simulate
at too coarse a lattice spacing, and the values used here are 
sufficiently fine so as to not approach the broken phase.
At these lattice spacings, it is clear that there is only a problem
if all three quarks are very near zero, which is not a physical point 
in the parameter space.

\begin{figure}
 \includegraphics[width=5in]{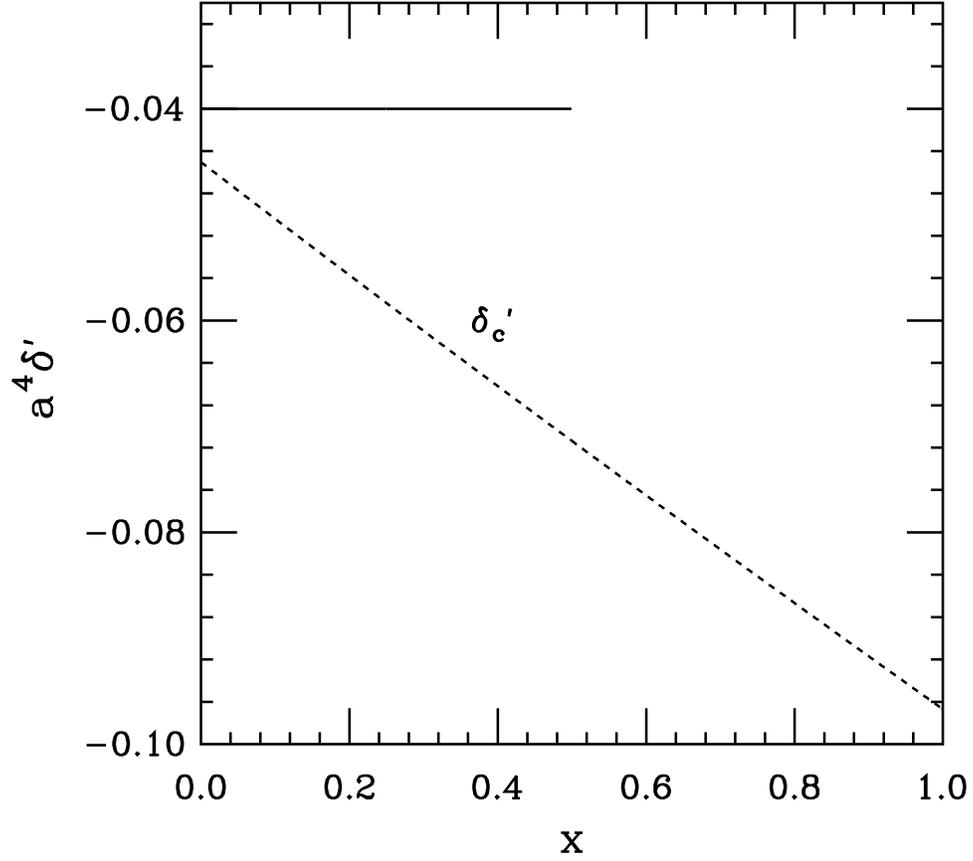} 
 \caption{$\delta'_c$ versus
   the fractional improvement of the splitting $\Delta_A$ for the coarse
   lattice. $x=1$ corresponds to the Asqtad action, while $x\approx 0.5$ 
   corresponds to HYP staggered fermions 
   \cite{Hasenfratz:2001hp,Hasenfratz:2002vv}. The horizontal solid
   line is the value of $\delta'_A$ for the Asqtad action for reference.
 }
 \label{fig:delta_vs_DA_c}
\end{figure}

One may wonder how this changes with more improved actions, such as
HYP staggered fermions \cite{Hasenfratz:2001hp}. 
Let us define $\tilde{\Delta}_A$ as the axial-taste splitting for some
more-improved action, and we define the quantity
\begin{equation}
  x \equiv \frac{\tilde{\Delta}_A}{\Delta_A},
\end{equation}
with $\Delta_A$ the Asqtad splitting.
We plot in
Fig.~\ref{fig:delta_vs_DA_c} for the coarse lattice spacing
$a^4\delta'_c$ as a function of $x$.
We only plot the coarse lattice here, since we
expect this to be the place where the problem would be more severe.
These are plotted with the lightest light quark mass from the MILC
simulations ($m_l/m_s=0.005/0.05$).  As $x$ is reduced, we see for
both the coarse and fine lattices that $\delta'_c$ is also reduced in
magnitude. For reference, HYP fermions lie roughly at $x\approx 0.5$
\cite{Hasenfratz:2002vv}.  Whether or not 
$|\delta'_A|$ is reduced by the same magnitude as $\Delta_A$ is
uncertain, and this is something that must be checked in a dynamical
simulation. A good improvement program should improve all the lattice
artifacts to some degree, so we would expect $|\delta'_A|$ to be
reduced as well. As can be see from the figure, as long as
$|\delta'_A|$ does not increase by a significant factor after
improvement (the central value for $\delta'_A$ in the Asqtad case
is shown by the solid line), then we will not
be in the broken phase.

\section{Conclusion}\label{sec:conc}

We have discussed the phase structure of a lattice theory with staggered
quarks, which has an approximate [to $\cO(a^2)$] $SO(4)$ taste symmetry,
as seen in the chiral theory. This taste symmetry can be spontaneously broken
down to $SO(3)$ for certain values of the parameters in the theory.
This most likely will not occur for current staggered actions, even 
among improved theories, unless all three quark masses are 
close to the chiral limit. At that point, there could be a phase transition
with 3 pseudo-Goldstone
bosons [with masses $\sim \cO(a^4)$], and a rather different
mass spectrum than in the unbroken phase.

The possibility of a flavor-symmetry (or in this case taste-symmetry)
breaking phase is a general feature of any lattice theory which has
additional 
terms in the
squared meson masses at finite lattice spacing. 
While originally discussed in the 
context of Wilson
fermions (Aoki phase), we see that there are complicated ways for the
taste symmetry with staggered fermions to be broken as well. 
The difference here is mainly
that the taste symmetry is already broken by the lattice spacing, and
it is only the approximate $SO(4)$ symmetry which is broken. So unlike
the Wilson case, there are significant corrections at higher order in
the lattice spacing.

This brings up an interesting question about the $\cO(a^4)$
corrections. We are stating here that when the lattice spacing becomes
too large, we are most likely to 
enter this broken phase. However, as $a$ increases,
higher-order scaling violations become more important. It
is unclear whether or not the broken phase will remain at extremely
coarse lattice spacing. Using a chiral Lagrangian that includes higher
order corrections \cite{Sharpe:2004is}, this
could be studied by determining the relative sizes of higher-order
contributions.

An interesting question arises relating to the quenched approximation
of staggered fermions. For Wilson fermions, there is the
possibility of entering the Aoki phase, and this can cause significant
problems with locality when simulating Wilson, Domain Wall, or Overlap
quarks \cite{Golterman:2003qe}. For the staggered case, however, there
is no Aoki phase in this context. 
There cannot be an Aoki phase in the context of the current
analysis, since the $\delta'_A$ splittings do not add to the mass of the
$\eta'_A$. So unlike the case of Wilson quarks, there is no
difficulty when simulating staggered quarks in the quenched
approximation.

\bigskip
\bigskip
\centerline{\bf ACKNOWLEDGMENTS}
\bigskip
We are grateful to 
M.\ Alford, 
C.\ Bernard, 
M.C.\ Ogilvie and
S.\ Sharpe
for helpful discussions, and especially C.\ Bernard for 
helpful comments on the manuscript.
This work was partially supported by the U.S. Department of Energy,
under grant number DE-FG02-91ER40628.

\vfill\eject

\end{document}